\documentclass[a4paper,11pt]{article}
\usepackage{pos}
\usepackage{epsfig}
\usepackage{natbib}
\usepackage{epstopdf}
\usepackage{mathrsfs}
\usepackage{slashed}
\usepackage{amsmath}
\usepackage{verbatim}
\usepackage{graphicx}
\usepackage{amssymb}
\usepackage{psfrag}
\usepackage{array}
\usepackage{subfig}
\newcommand{\ud}{\mathrm{d}}
\newcommand{\ue}{\mathrm{e}}

\newcommand{\bef}{\begin{figure}[htb]}
\newcommand{\eef}{\end{figure}}

\def\bea#1\eea{\begin{align}#1\end{align}}

\title{Heavy quark radiative energy loss in nuclei within the improved high-twist approach}

\author*[a]{Yi-Lun Du}

\affiliation[a]{Department of Physics and Technology, University of Bergen,\\ Postboks 7803, 5020 Bergen, Norway}

\emailAdd{Yilun.Du@uib.no}

\abstract{In this proceedings, we review our recent work on the heavy quark radiative energy loss in nuclei due to multiple parton scattering within the recently improved high-twist approach, where gauge invariance can be ensured by a delicate setup of the initial partons' transverse momenta. Our new result is consistent with the previous calculations of light quark energy loss in the massless limit and heavy quark energy loss in the soft gluon radiation limit, respectively. We show numerically the correction to the heavy quark energy loss as compared with previous result and with soft gluon radiation approximation. The necessity to go beyond soft gluon radiation limit is demonstrated for a global description of light and heavy flavor data in heavy-ion collisions.}

\FullConference{%
  HardProbes2020\\
  1-6 June 2020\\
  Austin, Texas}

\begin{document}
\maketitle

\section{Introduction}
\label{sec:intro} 
In the past decades, jet quenching phenomena have been intensively investigated via various observables in relativistic heavy ion collisions \cite{Gyulassy:2003mc,Majumder:2010qh,Qin:2015srf}. With the help of pQCD-based theoretical formalism for parton energy loss induced by multiple parton scattering in a nuclear medium, one can diagnose the fundamental properties of the cold nuclei and hot dense medium created in heavy-ion collisions by describing the experimental data on jet quenching \cite{Wang:2002ri,Bass:2008rv,Armesto:2009zi,Chen:2010te}. For example, the systematic extraction of jet quenching parameter, $\hat q$, has been carried out by the JET Collaboration \cite{Burke:2013yra}, based on a global fitting to the experimental data on single inclusive hadron production at RHIC and LHC within several jet quenching models. 

Among these jet quenching models, several are based on the high-twist expansion approach \cite{Guo:2000nz,Wang:2001ifa,Zhang:2003yn,Majumder:2009ge}, which encodes the effect of parton energy loss as higher twist corrections to the vacuum fragmentation functions. These models employ the generalized twist-4 factorization formalism as developed in Ref. \cite{Luo:1992fz, Luo:1994np}. The first calculation within this approach is performed in semi-inclusive electron-nucleus deep inelastic scattering (SIDIS) process \cite{Guo:2000nz, Wang:2001ifa}. Then it has been further improved by going beyond the helicity amplitude approximation \cite{Zhang:2003yn} and extended to evaluate heavy quark energy loss in SIDIS by considering charged-current interaction \cite{Zhang:2003wk, Zhang:2004qm}. However, to simplify these calculations, only a subset of Feynman diagrams at next-to-leading order (NLO) are considered by choosing appropriate physical gauge for the radiated gluon, which therefore doesn't allow for a consistent check of gauge invariance. 

The first complete NLO calculation at twist 4 has been carried out for the transverse momentum weighted differential cross section in SIDIS \cite{Kang:2013raa, Kang:2014ela}, and then extended to Drell-Yan lepton pair production in proton-nucleus collisions \cite{Kang:2016ron}. In these calculations, all Feynman diagrams with the nuclear size enhanced contributions are included. In the technical aspect, appropriate initial partons' transverse momenta flow has to be assigned to ensure the gauge invariance in the collinear expansion. In this proceedings, we apply this improved twist-expansion technique to revisit the final state quark energy loss in SIDIS. In particular, by focusing on the channel of charged-current interaction, it's allowed to evaluate light quark and heavy quark radiative energy loss on the same footing. We make comparison between our new results and the previous ones for both the light quark and heavy quark.  

\section{Heavy quark energy loss}
The calculation of heavy quark energy loss in SIDIS has already been performed in Ref.~\cite{Zhang:2004qm} where the transverse momenta for the rescattered gluons from the nucleus are set the same. All the relevant diagrams can be found in Refs.~\cite{Wang:2001ifa,Zhang:2004qm}. In order to obtain gauge invariant results at twist-4, the initial partons associated with the $2\to 2$ hard scattering have to satisfy on-shell conditions up to the leading order of transverse momenta of the rescattered gluons as specified in Ref.~\cite{Kang:2014ela,du2018revisiting}. This requirement can be fulfilled by a delicate setup for the transverse momenta of the initial partons from the nuclear target. 

Here we apply the same setup in the calculation of heavy quark energy loss in SIDIS \cite{du2018revisiting}. By considering the nuclear size enhanced contribution due to non-Abelian Laudau-Pomeranchuk-Migdal (LPM) interference in the twist-4 contributions and following the same ansatz as in Refs.~\cite{Zhang:2003wk,Zhang:2004qm} for the twist-4 parton matrix element, the averaged heavy quark energy loss, which is defined as the fraction of the energy carried by the radiated gluon, can be estimateed as  
\begin{equation}
\begin{aligned}
\langle\Delta z_g^Q\rangle(x_B, \mu^2)=&\int_0^{\mu^2}{\ud}\vec{\ell}_{T}^2\int_0^1{\ud}z \frac{\alpha_s}{2\pi}(1-z)\frac{\Delta\gamma_{Q\to Qg}(z,x_B,x_L,\vec{\ell}_{T}^2)}{\vec{\ell}_{T}^2+(1-z)^2M^2}\\
=&\frac{\mathrm{\tilde{C}C_A\alpha_s^2}x_B}{N_cQ^2x_A}\int_0^1{\ud}z\frac{1+(1-z)^2}{z(1-z)}\int_{\tilde{x}_M}^{\tilde{x}_\mu}{\ud}\tilde{x}_L\frac{(\tilde{x}_L-\tilde{x}_M)^2}{\tilde{x}^4_L}(1-{\ue}^{-\tilde{x}_{L}^2/x_A^2})a(z,M^2/\vec{\ell}^2_T),
\label{eq-deltaz}
\end{aligned}
\end{equation}

\begin{equation}
\begin{aligned}
a(z,M^2/\vec{\ell}^2_T)=&\frac{(1+z)}{2}-\frac{2(1-z)^3z(1+z)}{1+z^2}\frac{M^2}{\vec{\ell}_T^2}+\frac{(1-z)^4(3z^3-5z^2+7z-1)}{2(1+z^2)}\frac{M^4}{\vec{\ell}_T^4}\\
&+\mathrm{\frac{2C_F}{C_A}}\left[(1+z)^2+(1-z)^4\frac{M^2}{\vec{\ell}_T^2}\right]\frac{(1-z)^4}{1+z^2}\frac{M^2}{\vec{\ell}_T^2}.
\label{eq-a1}
\end{aligned}
\end{equation}
The coefficient $\mathrm{\tilde{C}}$ is proportional to the gluon distribution inside a nucleon, and the suppression factor $1-{\ue}^{-\tilde{x}_L^2/x_A^2}$ is due to the LPM interference. The main result in this work $a(z,M^2/\vec{\ell}^2_T)$ differs from that of Ref.~\cite{Zhang:2004qm} by
\bea
\Delta a\equiv a-a^{\text{Ref.~\cite{Zhang:2004qm}}}
=\left [z-\frac{1}{2}+\mathrm{\frac{C_F}{C_A}}(1-z)^2\right ]\left [(1+z)^2+(1-z)^4\frac{M^2}{\vec{\ell}_T^2}\right ]\frac{(1-z)^2}{1+z^2}\frac{M^2}{\vec{\ell}_T^2}.
\label{eq-diff-a}
\eea
One can see that this new correction is proportional to $(1-z)^2M^2$, which vanishes in the massless limit $M\to 0$ or soft gluon radiation limit $z\to 1$. In massless limit, $a$ in Eq.(\ref{eq-a1}) will reduce to $(1+z)/2$, which reproduces the previous result in Ref.~\cite{Zhang:2003yn}. That is to say, the final result of light quark energy loss in Refs.~\cite{Guo:2000nz,Wang:2001ifa,Zhang:2003yn} is gauge invariant, and thus its phenomenological studies remain valid. On the other hand, the final
result of heavy quark energy loss in Refs.~\cite{Zhang:2003wk,Zhang:2004qm} is complete and gauge invariant only in the soft gluon radiation limit, which has been employed in phenomenological application of heavy flavor production in heavy-ion collisions \cite{cao2013heavy,PhysRevC.92.024907,cao2018heavy}. For future complete phenomenological investigations of heavy quark energy loss beyond soft gluon limit, we should instead employ the our new gauge invariant result.  

\section{Numerical result}

\begin{figure*}[tb]
\centering
\subfloat[][\empty]{
\begin{minipage}[t]{0.48\linewidth} 
\centering 
\includegraphics[width=2.8in]{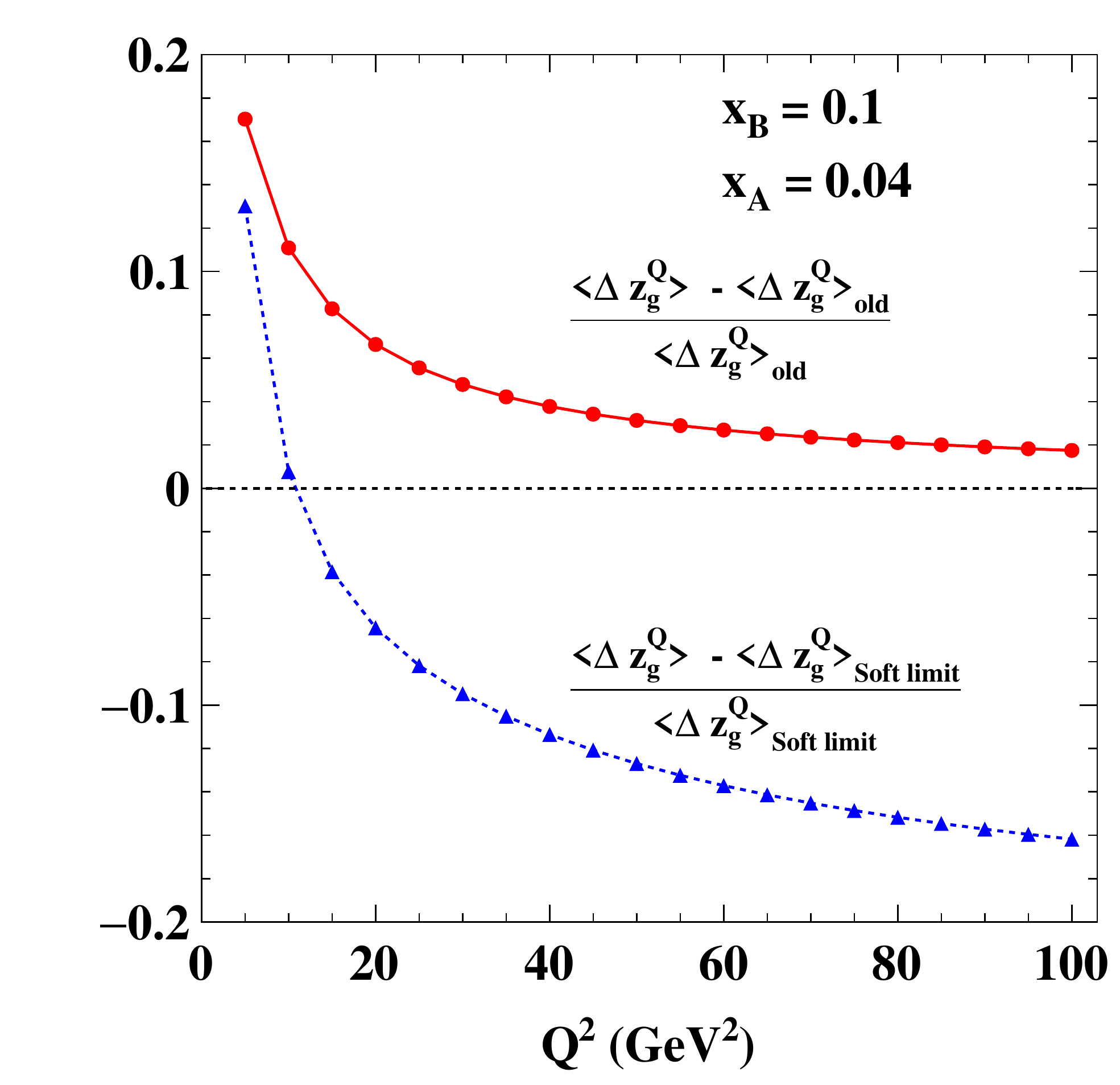} 
\end{minipage}%
} 
\subfloat[][\empty]{
\begin{minipage}[t]{0.48\linewidth} 
\centering 
\includegraphics[width=2.8in]{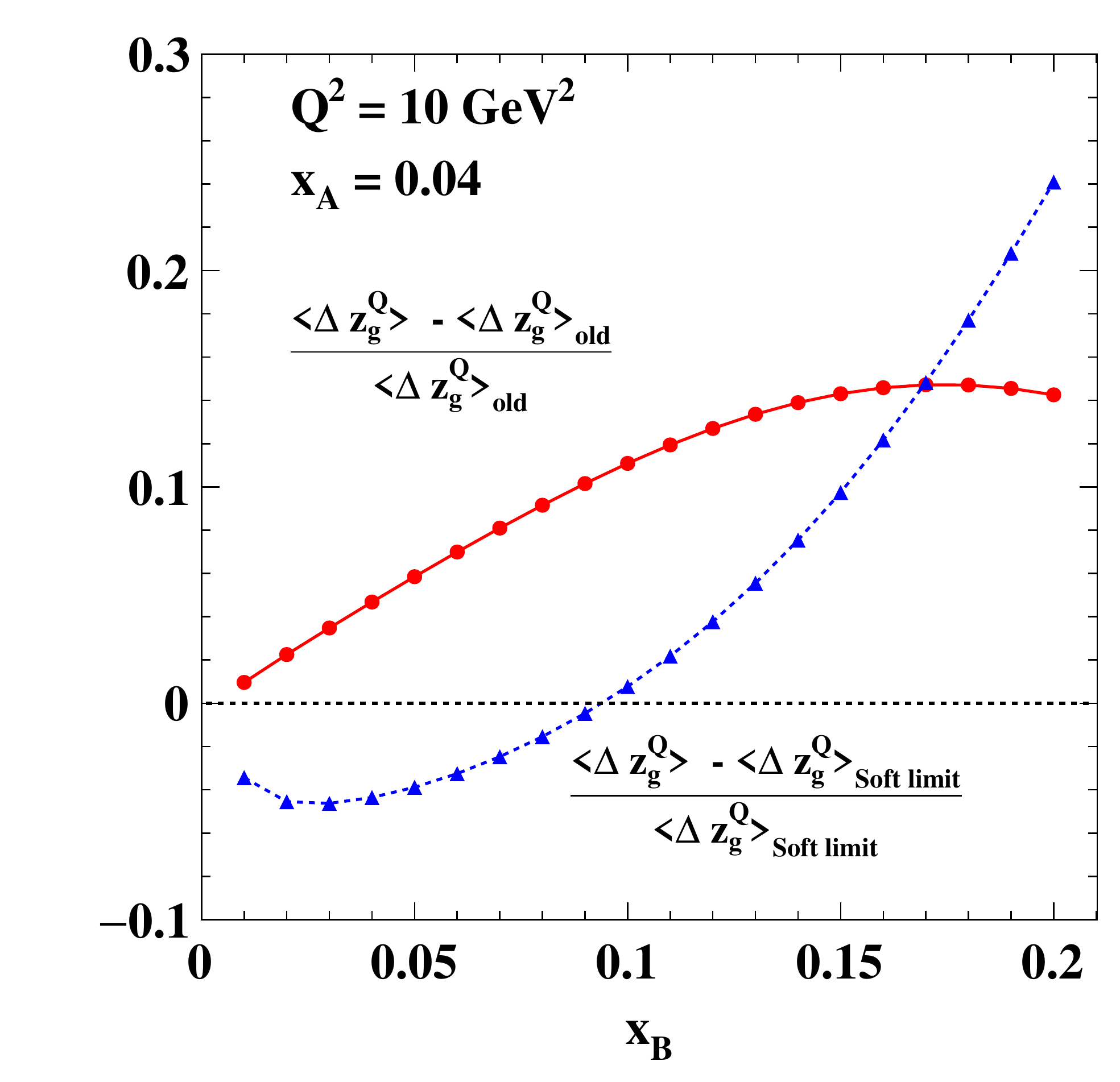} 
\end{minipage} 
}\caption{ The $Q^2$ and $x_B$ dependence of the relative correction of charm quark energy loss $\delta\langle\Delta z_g^Q\rangle/\langle\Delta z_g^Q\rangle$ as compared to that from old work~\cite{Zhang:2004qm} (red solid) and with soft gluon approximation (blue dashed).}
\label{fig-DELTAE}
\end{figure*}

\begin{figure*}[bt] 
\centering
\subfloat[][\empty]{
\begin{minipage}[t]{0.48\linewidth} 
\centering 
\includegraphics[width=2.8in]{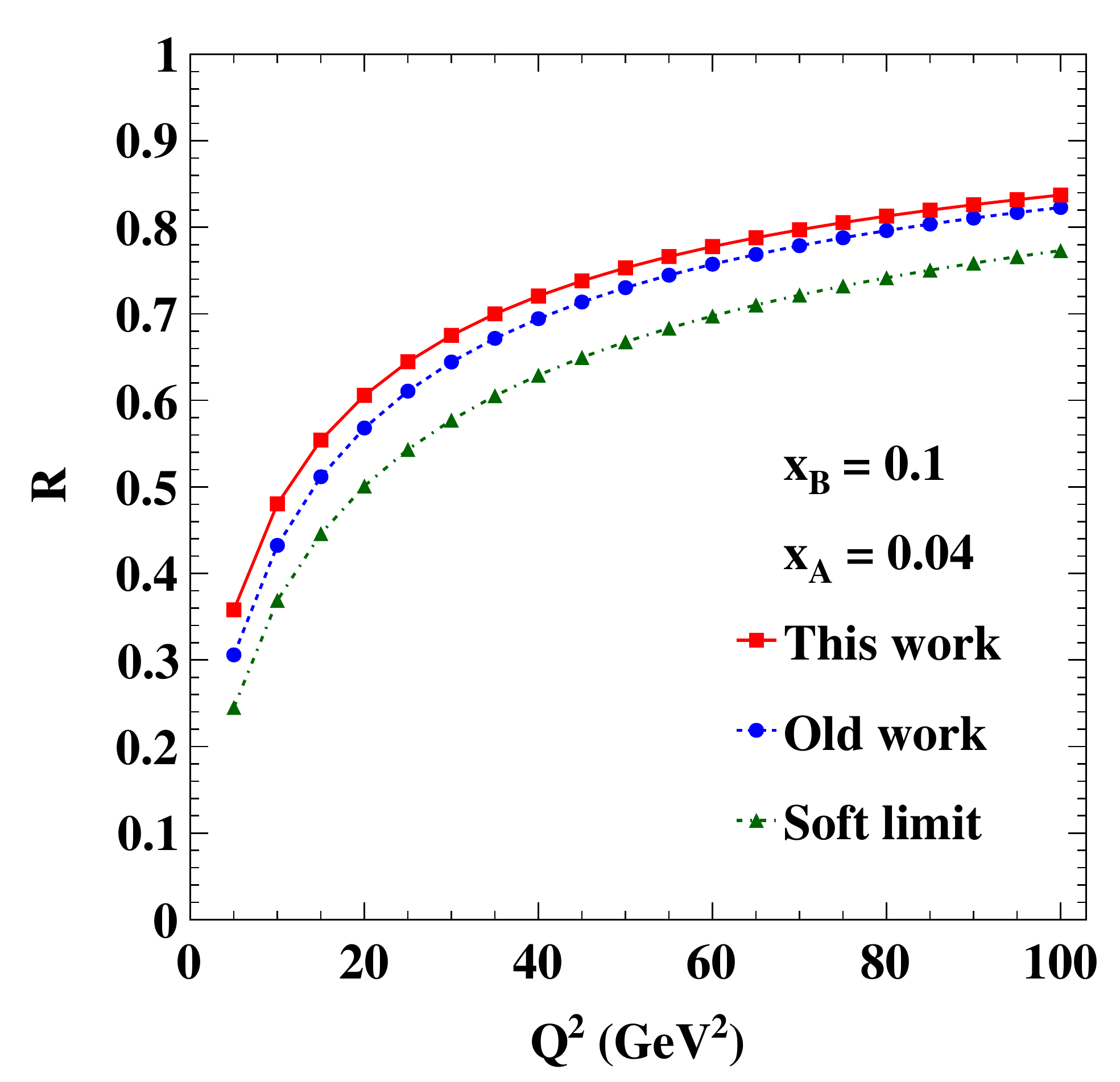} 
\end{minipage}%
} 
\subfloat[][\empty]{
\begin{minipage}[t]{0.48\linewidth} 
\centering 
\includegraphics[width=2.8in]{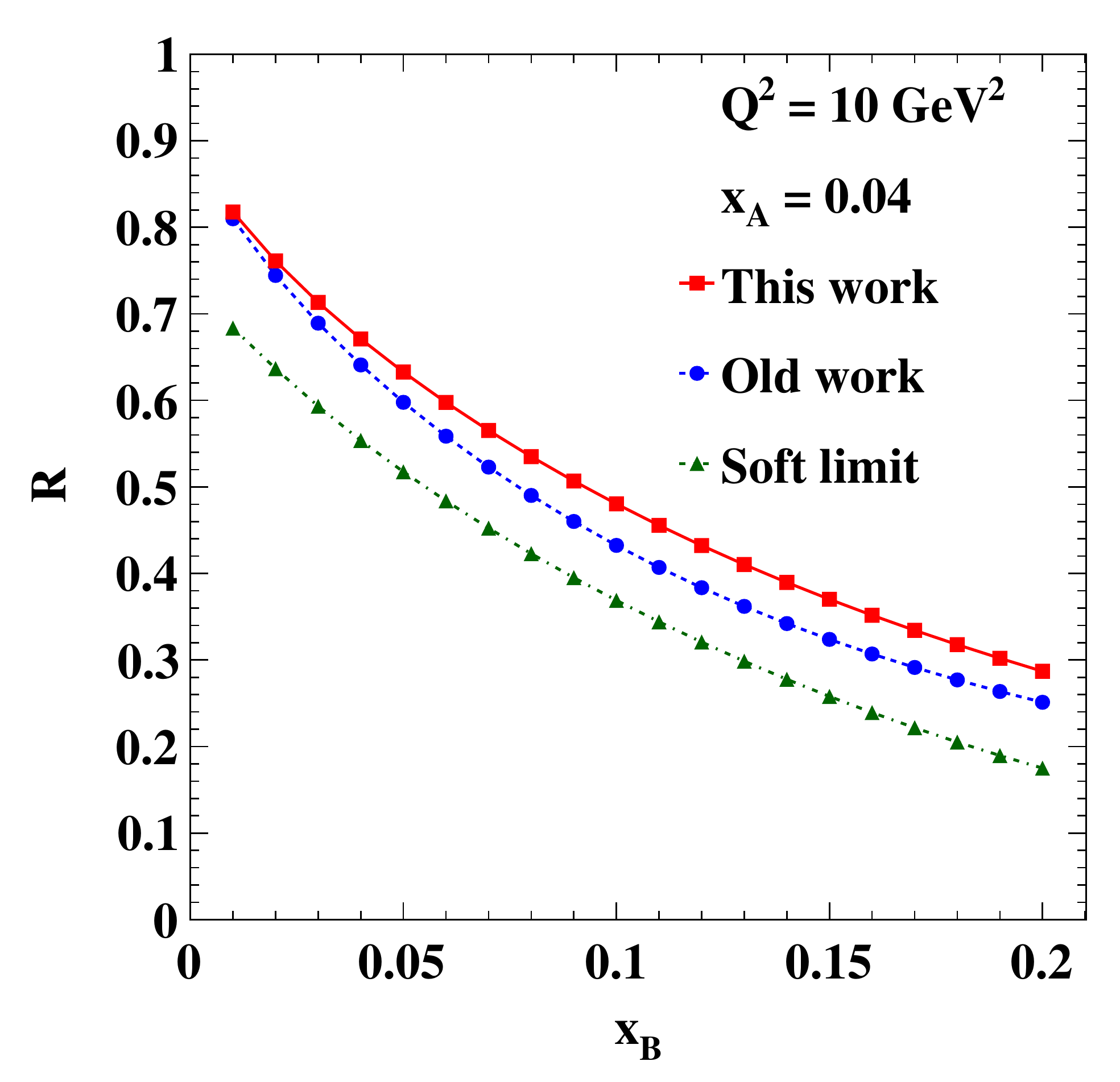} 
\end{minipage} 
}\caption{The $Q^2$ and $x_B$ dependence of the ratio $R$ between charm quark and light quark energy loss from this work (red solid), old work~\cite{Zhang:2004qm} (blue dashed), and soft limit (green dashed-dotted).}
\label{fig-R}
\end{figure*} 

To evaluate the new correction to heavy quark energy loss numerically, we choose charm quark mass $M = 1.5$ GeV and $x_A = 0.04$ for a nucleus with a radius $R_A=5$ fm. In the left panel of Fig.~\ref{fig-DELTAE}, we show the relative correction between our new result of heavy quark energy loss and that in Ref.~\cite{Zhang:2004qm} as a function of $Q^2$ with fixed $x_B=0.1$ by the red solid curve. One can see that the new correction leads to significant extra heavy quark energy loss in the small $Q^2$ region. However, the correction becomes negligible in the large $Q^2$ region. It is understandable from Eq.~(\ref{eq-diff-a}) since the correction is proportional to $M^2$, and thus power suppressed in the large $Q^2$ region as other higher-twist effects. The correction between our new result and the soft gluon radiation limit is shown by the blue dashed curve, which can be as large as $16\%$ in the large $Q^2$ region. Therefore it is necessary to consider contributions beyond the soft limit in future more precise phenomenological studies of heavy quark energy loss in heavy-ion collisions. The right panel of Fig.~\ref{fig-DELTAE} shows the relative correction between our new results and the previous one in Ref.~\cite{Zhang:2004qm} as a function of $x_B$ with fixed $Q^2=10$ GeV$^2$ by the red solid curve. The correction becomes significant for small initial heavy quark energy (large $x_B$) as shown by the red solid curve. The contribution beyond the soft gluon limit is also appreciable for small initial heavy quark energy (large $x_B$) as shown by the blue dashed curve.

We show the ratio of charm quark and light quark energy loss $R=\langle\Delta z_g^Q\rangle(x_B,\mu^2)/\langle\Delta z_g^q\rangle(x_B,\mu^2)$ in Fig.~\ref{fig-R}. The light quark energy loss $\langle\Delta z_g^q\rangle(x_B,\mu^2)$ can be easily obtained by taking massless limit $M\to 0$ in Eq. (\ref{eq-deltaz}). In Fig.~\ref{fig-R}, we show the dependence of $R$ on $Q^2$ with fixed $x_B=0.1$ in the left panel
and that on $x_B$ with fixed $Q^2=10$ GeV$^2$ in the right panel. One can observe the reduction of energy loss of heavy quark compared with that of light quark ($R<1$) due to the effect of the dead cone in our new result (red solid curve) and the previous one ~\cite{Zhang:2004qm} (blue dashed curve). Such a reduction is due to the heavy quark mass and therefore should disappear at high $Q^2$ and large initial quark energy (small $x_B$). For comparison, we show $R$ with soft gluon radiation limit taken in both light and heavy quark energy loss at the same time (the green dashed-dotted curve). One can see the significant contributions from beyond the soft gluon limit by comparing the red solid curve and the green dashed-dotted curve.

\section{Summary}
\label{sec-sum}
In this proceedings, we revisited a series of studies~\cite{Guo:2000nz,Wang:2001ifa,Zhang:2003yn,Zhang:2003wk,Zhang:2004qm} on quark radiative energy loss in SIDIS off a nuclear target due to multiple parton scattering with the improved twist-4 factorization formalism where gauge invariance is ensured. We found that the result of light quark energy loss remains the same. But for heavy quark energy loss, our new result leads to correction if beyond the soft gluon limit. To demonstrate the significance of this correction quantitatively, we evaluated numerically the heavy quark energy loss and compared with the previous one in Ref.~\cite{Zhang:2004qm}. We found noticeable correction in the small $Q^2$ and large $x_B$ regions. Our new result was also compared with that with soft gluon approximation. The significant corrections with respect to these two suggest the necessity of employing the complete and gauge invariant result (beyond soft limit) for more precise description of heavy flavor jet quenching in heavy-ion collisions. This also has phenomenological impact on precise extraction of the jet transport coefficient from light and heavy flavor data in heavy-ion collisions. 

\acknowledgments
This proceedings is based on the work in collaboration with Yayun He, Xin-Nian Wang, Hongxi Xing and Hong-Shi Zong. I acknowledge the support from National Natural Science Foundation of China under Grants No. 11475085, No. 11535005, No. 11690030, and National Major state Basic Research and Development of China under Grants No. 2016YFE0129300; and Horst Stoecker and the support from the Helmholtz Graduate School HIRe for FAIR, the F\&E Programme of GSI Helmholtz Zentrum f$\ddot{\mathrm{u}}$r Schwerionenforschung GmbH, Darmstadt, the Giersch Science Center, the Walter Greiner Gesellschaft zur F$\ddot{\mathrm{o}}$rderung der physikalischen Grundlagenforschung e.V., Frankfurt, and the AI grant of SAMSON AG, Frankfurt. 



\bibliographystyle{apsrev4-1}
\bibliography{biblio}

\end{document}